\documentstyle[11pt,paspconf]{article}

\begin{document}
\title{A Large Proper-Motion Survey in Plaut's Low-Extinction Window
\altaffilmark{1}}
\author{R. A. M\'endez}
\affil{European Southern Observatory, Karl-Schwarzschild Stra$\beta$e 2,
D-85748, Garching b. M\"unchen, GERMANY}
\author{R. M. Rich\altaffilmark{2}}
\affil{Columbia University, Astronomy Department, 538 W. 120th St. Box
42 Pupin, New York, NY 10027, USA}
\author{W. F. van Altena and T. M. Girard}
\affil{Yale University, Astronomy Department, P. O. Box 208101, New Haven, CT
06520-8101, USA}
\author{S. van den Bergh}
\affil{Dominion Astrophysical Observatory, 5071 W. Saanich Rd.,
Victoria, BC, V8X 4M6, CANADA}
\author{S. R. Majewski}
\affil{University of Virginia, Department of Astronomy, P. O. Box
3818, Charlottesville, VA 22903-0818, USA, and Carnegie Observatories,
813 Santa Barbara Street, Pasadena, CA 91101, USA}
\altaffiltext{1}{To appear in 4th ESO/CTIO Workshop on THE GALACTIC
CENTER, La Serena, Chile, 10--15 March 1996, published by the
Astronomical Society of the Pacific, ASP Conference Series}
\altaffiltext{2}{Visiting Astronomer, Cerro Tololo Inter-American
Observatory which is operated by AURA, Inc.\ under cooperative
agreement with the National Science Foundation} 

\begin{abstract}
We present preliminary results from the deepest and largest
photographic proper-motion
survey ever undertaken of the Galactic bulge. Our first-epoch plate
material (from 1972-3) goes deep enough ($V_{lim} \sim 22$)
to reach below the bulge main-sequence turnoff. These plates cover an
area of approximately $25' \times  25'$ of the bulge in the
low-extinction ($A_v \sim 0.8$ mag) Plaut
field at $l= 0\deg, b= -8\deg$, approximately 1 kpc south of the
nucleus. This is the point at which
the transition between bulge and halo populations likely occurs and is,
therefore, an excellent location to study the interface between the
dense metal-rich bulge and the metal-poor halo.

In this conference we report results based on three first-epoch
and three second-epoch plates spanning 21 years. It is found that it 
is possible to
obtain proper-motions with errors less than 0.5 mas/yr for a
substantial number of stars down to V= 20, without color restriction.
For the subsample with errors less than 1 mas/yr we derive
proper-motion dispersions in the direction of Galactic longitude and latitude of 
$3.378 \pm 0.033$ mas/yr and $2.778 \pm 0.028$ mas/yr
respectively. These dispersions agree with those derived by 
Spaenhauer et al. (1992) in Baade's window.
\end{abstract}
\keywords{Galactic bulge, kinematics, proper-motions, orbits, bar}
\section{Introduction}
Historically, tremendous observational effort has been invested in
understanding the formation of the halo and chemical evolution of the
disk. Ideally, one wants to measure proper-motions, radial velocities,
and abundances for members of a stellar population, inspired
by the seminal effort of Eggen, Lynden-Bell \& Sandage (1962).
While the halo does not appear to exhibit clear correlations
of abundances with kinematics (cf. Carney et al. 1990), there is some
indication that metal-rich bulge stars have a smaller velocity dispersion
(Rich 1990). In much larger samples of bulge giants in fields
away from the minor-axis, Minniti (1993, 1996) finds abundance/kinematics
trends that suggest that more metal-rich bulge giants have greater
rotational support. Unfortunately, fields distant from the minor
axis may be contaminated with giants at the tangent point
arising from the disk population (Tiede and Terndrup 1996). 
The next step in confirming
Minniti's findings is the careful study of a minor-axis field, including
proper-motions.   

These issues can
be settled by correlating abundances with proper-motion and radial
velocity dispersions. The geometry for viewing the bulge is favorable:
we are 8.5 kpc distant from a system which has most of the mass contained
within 1 kpc.  We are therefore privileged to study the bulge from
an almost extragalactic perspective.

Given its extreme crowding, high extinction, and southerly
declination, the bulge has received substantially less observational
attention than the globular cluster system.
Only one proper-motion study has been undertaken (Spaenhauer
et al. 1992).  While a landmark achievement, their work addresses the
extremely crowded Baade's Window field $(l=0\deg, b=-4\deg)$. There
were three first-epoch plates, all obtained on the Palomar 200-inch
telescope. Because the plates
were B plates, very few late M giants were measured; these stars
dominate the bulge asymptotic giant branch, and it is of great
interest to compare their kinematics with bulge K giants.

Our study addresses the tangential kinematics of the Galactic
bulge in Plaut's (Plaut 1970, 1971)
low-extinction window ($E_{B-V}= 0.25$ mag, van den Bergh \& Herbst 1974). This
window, centered at $l=0\deg, b=-8\deg$, provides a unique place to look at the Galactic
bulge and its transition into the the halo of the Galaxy, as the line-of-sight
crosses the Galactic minor-axis some 1.3 kpc to the South of the nucleus.
Furthermore, this region has smaller reddening and is less crowded than
Baade's window ($E_{B-V}=0.42$ mag, Blanco and Blanco 1985).

The field in this study is of critical importance as it lies at the
edge of the bulge (as defined, for example, by the {\sl COBE-DIRBE\/} map).
It has extinction $A_v= 0.8$ mag, lower than most bulge fields, and it will
contain a substantial number of halo giants, allowing one to probe the
bulge/halo transition. We know little about the field population of
the inner halo; this study will also fill that gap in our knowledge.

We expect to measure CCD photometry in the B, V, and I passbands, 
and proper-motions
for an unbiased sample of approximately 30,000 stars in our minor-axis field.  
We hope to further obtain radial velocities and low resolution
abundances for about 5,000 stars.
A large, unbiased sample is important because much of the 
outcome depends on dividing the data into
subsamples as a function of abundance or kinematics.
\section{Plate material, measurements, and the proper-motions}
Our project is based on a unique sample of twenty photographic plates
of a Galactic bulge field on the minor-axis at $b= -8\deg$
obtained by Sidney van den Bergh in 1972-3 using the Kitt Peak
84-inch and the 200-inch telescopes (van den
Bergh \& Herbst 1974). The 100-inch telescope at Las Campanas 
has been used to obtain thirteen second epoch plates in 1993; deep
intermediate epoch plates of this field (1979) were obtained by Jeremy
Mould also at Las Campanas (these intermediate epoch plates have not
been used in this preliminary study).

The plates were digitized using the the Yale PDS 2020G laser
interferometer/microdensitometer measuring machine in raster-scan
mode. The aperture size was $33 \; \mu m$, while the step size was
$30\; \mu
m$. At this step size, the faintest images will be properly sampled
for the astrometric centroiding (our plates have scales on the order
of 10 arc-sec/mm). All
scans were performed under the best
possible thermal conditions to avoid instrumental drifts during the
scan. Each scan took between 8 and 10 hours. Depending on the plate
scale, the digitized frames had between $4,200 \times 4,200 \; pixels^2$ and
$5,600 \times 5,600 \; pixels^2$. The PDS system outputs an integer fits-file
with 16 bits/pixel that is easily converted into other formats for later
analysis. At this conference we report on preliminary results from
reductions of only three first-epoch and three second-epoch plates.

Analysis of the Yale-PDS microdensitometer data routinely yields star 
centroids
to 1/20 of a pixel (20 mas).  A star measured on five plates in each
color will have its position known at least 2--3 times
better than this, and over the 21 yr baseline we expect errors no
larger than 0.5 mas/yr in each color.  
This corresponds to approximately 20 km/s at the distance of the
Galactic Center. This is comparable to Spaenhauer et al., and
matches the accuracy with which radial velocities in our spectroscopic
follow-up will be measured.

The first step in the process of obtaining proper-motions was to
create an input
catalogue of approximate stellar positions. For this purpose we 
selected a relatively deep second-epoch master plate with the best image and fog characteristics,
requiring also that the full FOV would be available in this master
plate. Then, we created a master list of candidates to perform the astrometric
solution. This was done by running
DAOFIND from the DAOPHOT package within IRAF. Approximate
``image'' parameters (FWHM) and ``frame'' parameters (the rms variation on the plate fog) were
computed from the digitized frame of the master plate. The master
list contains a little less than 100,000 detections at an $8 \sigma$
level. Visual inspection of the detections against the digitized plate
confirmed that no obvious stars will be left-out of the master list,
even in cases of relatively high crowding.

This master list was then used to ``feed'' the Yale Image Centering
routine described by Lee \& van Altena (1983), which provides image
centers, instrumental magnitudes and centering error estimates.

The next step was to reduce all of the plates to the master plate.
No preliminary corrections for atmospheric refraction or distortion were
performed in this solution. Refraction is likely to be an important
effect, particularly
for the first-epoch plates. However, this effect would be absorbed in
the plate constants when converting to the master plate. The
centered list on each plate was converted to the master plate using
interactive software designed to recognize and isolate outliers, and
to include distortion terms also in an interactive way until the
residuals show no systematic trends.

The plate transformations described above indicate that we are indeed
able to reach the expected level of accuracy. For example, plate-to-plate
transformations between pairs of same-epoch plates have an rms scatter
of 35 mas (i.e., 25 mas on each plate). On the other hand,
transformations between first and second epoch plates indicate an rms
dispersion 
of 80 to 100 mas. This higher dispersion is due to the proper-motion dispersion of the
stars: For a population of velocity dispersion of 120 km/sec
at 8 kpc, we expect an rms dispersion of 75 mas (21 yr baseline).
Convolving the 75 mas with our measurement error gives our measured
80 mas dispersion. Evidently, the full analysis (including a
pre-correction for optical distortion and atmospheric refraction) will
give more precise results, but these preliminary reductions show that
we can achieve the required accuracy from the available plate material.
\begin{figure}
\vspace{6.8cm}
\caption{Proper-motion error {\it vs.} V magnitude. Upper panel is for
Galactic longitude, lower panel is for Galactic latitude. Note the
rapid increase in the errors for $V> 18$.}
\end{figure}
Once all the plates had been transformed to the master plate, and
distortions had been removed, we computed the proper-motions by
performing a {\it weighted} least-square linear fit to position {\it
vs}. time, taking
into account the individual plate solutions. This solution gave
not only proper-motions but
also error estimates for the derived motions. The individual proper-motion
errors were computed from the scatter of the residuals about the best-fit line
as well as from the formal error of the slope in the linear fit.
Figure 1 shows our proper-motion errors in Galactic longitude ($\mu_l$) and
latitude ($\mu_b$) as a function of apparent V magnitude. For $V >
18$ the errors
increase sharply. Also, we have very few objects with $V >
20$, mainly because of the magnitude limit for the 1st-epoch 
84-inch plates. The use of the 200-inch plates will allow us to go
to $V \approx 22$ for a selected sample of stars. Table 1 indicates
the mean proper-motion errors in $\mu_l$ and $\mu_b$ as
a function of V magnitude, as well as the number of stars
in each magnitude interval. For $V > 18$ the errors
become of the same order as the 
expected proper-motion dispersion ($\Sigma_{\mu} \approx 3 \;
mas/yr$). At this magnitude (or fainter), the sample with 
proper-motion errors less than about 1 mas/yr will only 
contain objects far from the mean error, and their errors are 
probably not very well determined. This 
translates into a spuriously large proper-motion dispersion for the
fainter objects (see Table 2).
\begin{table}
\caption{Proper-motion error {\it vs.} V magnitude.} \label{tbl-2}
\begin{center}
\begin{tabular}{cccc}
\tableline
\tableline
Magnitude range& Number of stars & $\sigma_{\mu_l}$ &
$\sigma_{\mu_b}$\\
 & & mas/yr & mas/yr\\
\tableline
$14.0 \le V < 16.0$ & 1530 & 0.75 & 0.69 \\
$16.0 \le V < 17.0$ & 3944 & 0.91 & 0.88 \\
$17.0 \le V < 17.5$ & 3023 & 1.47 & 1.44 \\
$17.5 \le V < 18.0$ & 4102 & 2.18 & 2.13 \\
$18.0 \le V < 18.5$ & 5904 & 3.12 & 3.05 \\
$18.5 \le V < 19.0$ & 6392 & 4.10 & 4.00 \\
$19.0 \le V < 22.0$ & 6287 & 5.32 & 5.23 \\
\tableline
\tableline
\end{tabular}
\end{center}
\end{table}
\section{Analysis of the proper-motions}
The vector-point diagram for proper-motion error cuts at 1 mas/yr is 
shown in Figure 2. This figure compares well with Figure 1 in 
Spaenhauer et al. (1992). This suggests that the
proper-motion dispersion for these two fields is approximately the same,
as it is indeed the case (Table 2).
\begin{figure}
\vspace{6.7cm}
\caption{Vector-point diagram for objects with errors less than 1
mas/yr in each coordinate. This Figure compares very well with 
Figure 1 in Spaenhauer et al. (1992)}
\end{figure}
The combination of a large sample and small measurement errors means
that the proper-motion dispersions are known with great
accuracy. Following Spaenhauer et al. (1992) the true (error-corrected)
proper-motion dispersion, $\Sigma_\mu$, is given by:
\begin{equation}
\Sigma_\mu^2 = {1\over (n-1)}\sum_{i=1}^n(\mu_i - \bar\mu)^2 -
{1\over n} \sum_{i=1}^n\sigma_{\mu_i}^2
\end{equation}
where $n$ is the sample size, $\mu$ is one component of the proper-motion,
and $\sigma_{\mu_i}$ is the error of a single proper-motion measurement. The
error in $\Sigma_\mu$ in any subsample of $n$ stars is given by:
\begin{equation}
\xi_\Sigma = \left( {1\over 2n}\Sigma_\mu^2 +
{1\over 2n^2\Sigma_\mu^2}\sum_{i=1}^n
{\sigma_{\mu_i}^4\over n_i}\right)^{1/2}
\end{equation}
where $n_i$ is the number of plates on which star $i$ is measured. Spaenhauer et al. (1992) measured proper-motion dispersions
of approximately 3 mas/yr. The above equations show that for a typical error of 0.5 mas/yr
for a single
measurement (Spaenhauer et al.'s level) our results would be unaffected by
measurement errors. For subsamples of 200 stars, we estimate that the error in
each subpopulation dispersion would be approximately 0.1 mas/yr (4
km/s at the distance of the Galactic bulge).

We have computed (intrinsic) proper-motion dispersions along Galactic 
longitude ($\Sigma_{\mu_l}$) and latitude ($\Sigma_{\mu_b}$) using 
Equations (1) and (2). In order to properly handle outliers, dispersions were
determined in an iterative way following a procedure similar to the
technique of outlier elimination using probability plots (M\'endez \&
van Altena 1996). The results for the dispersions are shown in Table 2.

Table 2 shows how important it is to have a large sample of
small-error proper-motions; the dispersions are determined with very
high accuracy, typically the errors are less than 2\%. Also, it can be
seen that the proper-motion dispersion {\it does not} seem to change with
apparent magnitude within the uncertainties 
(except for the large-error bin at $18 \le V < 22$). There may be an indication that
$\Sigma_{\mu_l}$ is slightly increasing with apparent magnitude, but the
change is only at the $2 \sigma$ level. Since we {\it do} expect to see a
mixture of populations along the line-of-sight, each with different
dispersion and mean rotational motion, this result implies that the
contamination by these populations is rather minimal. 
The stellar ratio of disk/bulge, thick-disk/bulge, and halo/bulge is
expected to change as a function of V magnitude, and so, therefore, 
is the 
proper-motion dispersion. We do not expect the disk or thick-disk to
make a very
large contribution (e.g., at Baade's window their contribution is less
than 20\%, and our field is at twice the Galactic latitude). However,
we {\it do} expect to have a rather significant contribution from stars
in the inner halo, which we do not seem to detect.
\begin{table}
\caption{Intrinsic proper-motion dispersion {\it vs.} V magnitude 
(only stars with errors less than 1 mas/yr in each coordinate included
in solutions).}\label {tbl-3}
\small
\begin{center}
\begin{tabular}{ccccc}
\tableline
\tableline
Magnitude range& No. stars in l & $\Sigma_{\mu_l}$ & No. stars in b & 
$\Sigma_{\mu_b}$\\
 & & mas/yr & & mas/yr\\
\tableline
$14.0 \le V < 16.0$ & 1136 & $3.279 \pm 0.069$ & 1136 &  $2.811 \pm 0.059$\\
$16.0 \le V < 16.5$ & 1442 & $3.272 \pm 0.061$ & 1438 &  $2.872 \pm 0.053$\\
$16.5 \le V < 17.0$ & 1211 & $3.432 \pm 0.070$ & 1206 &  $2.674 \pm 0.055$\\
$17.0 \le V < 18.0$ & 1299 & $3.464 \pm 0.068$ & 1293 &  $2.748 \pm 0.054$\\
$18.0 \le V < 22.0$ &  236 & $5.293 \pm 0.244$ &  238 &  $5.270 \pm 0.240$\\
\\
$14.0 \le V < 18.0$ & 5088 & $3.378 \pm 0.033$ & 5077 &  $2.778 \pm 0.028$\\
Spaenhauer et al.   &  429 & $3.2   \pm 0.1$   &  429 &  $2.8   \pm 0.1$\\
\tableline
\tableline
\end{tabular}
\end{center}
\end{table}
Table 2 also shows that the proper-motion dispersions $\Sigma_{\mu_l}$
and $\Sigma_{\mu_b}$ {\it are}
different in Galactic latitude and longitude at a $14 \sigma$ level,
which is a much more definitive result than that of Spaenhauer et al.'s (shown on
the last line of Table 2), who suggested differences at the $3
\sigma$ level. On the other hand, our results do agree
with Spaenhauer et al.'s results within their rather large uncertainties in
both $\Sigma_{\mu_l}$ and $\Sigma_{\mu_b}$.

The interpretation of the values listed in Table 2 in terms of velocity
dispersions for the bulge stars is complicated due to the expected
contamination from halo stars. A simple two-component model predicts
that the velocity dispersions for bulge stars in Galactic longitude and
latitude ($\Sigma_{B_{l}}$ and $\Sigma_{B_{b}}$ respectively) are given by:
\begin{equation}
\Sigma_{B_{l}}^2 = (1+x) \Sigma_l^2 -x \Sigma_{H_{l}}^2 - 
\frac{x}{1+x}<V_B>^2
\end{equation}
and
\begin{equation}
\Sigma_{B_{b}}^2 = (1+x) \Sigma_b^2 -x \Sigma_{H_{b}}^2
\end{equation}
where $\Sigma_l$ and  $\Sigma_b$ are the total velocity dispersions
in Galactic longitude and latitude respectively, $\Sigma_{H_{l}}$ and
$\Sigma_{H_{b}}$ are the halo velocity dispersions in Galactic
longitude and latitude, $<V_B>$ is the mean rotation for the
bulge, and $x$ is the ratio of the number of halo to bulge stars
in our sample.

Assuming 8.5 kpc as the distance to the Galactic center, and the
mean values derived from Table 2, Equations (3) and (4) imply that
$\Sigma_{B_{l}} \approx 140 \; km/s$ and $\Sigma_{B_{b}} \approx 120
\; km/s$. The larger $\Sigma_{B_{l}}$ is consistent with rotation
broadening and anisotropy of the Galactic bar (Zhao 1996, Zhao et al.
1996), but it does not necessarily argue for triaxiality. Zhao et al.
(1994) have shown that only a strong {\it vertex deviation} of the
bulge velocity ellipsoid (i.e., a non-zero cross-term $<V_rV_l>$) will
be a definitive indication of triaxiality.
\section{Conclusions and the future}
We can obtain proper-motions for a large sample of
bulge stars with errors small enough to allow a meaningful
kinematical study of the bulge. It is found that the proper-motion
dispersion in our field is comparable with that found by Spaenhauer et
al. (1992) in Baade's window. Our dispersions are consistent with
broadening by rotation and anisotropy of the Galactic bar as predicted
by the dynamical models of Zhao et al. (1996).

Radial velocities from our spectroscopic survey will be extremely important
as a complement to the proper-motions to confirm the presence
of a bar. Triaxiality will take the form of a vertex deviation in the 
$\Sigma_r$ vs. $\Sigma_l$ plane, as suggested from the Spaenhauer et
al. data analysed by Zhao et al. (1994), and from a larger
spectroscopic follow-up of Spaenhauer et al.'s sample by Rich et al. (1996).

If the bulge collapsed and spun up as metallicity increased, we should
see $\Sigma_r$ and $\Sigma_b$ decrease with higher metallicity
(Minniti 1993, 1996). In a rapidly rotating population, integration
through the line of sight will reveal that $\Sigma_l$ will be
artificially broadened (Zhao et al. 1994, 1996). Applied to our large
minor-axis sample this analysis will help constrain the
formation/enrichment history of the bulge.
\acknowledgments
We are grateful to Jeremy Mould for lending us the deep
intermediate-epoch plates taken by him in 1979. We are also grateful
to Kyle Cudworth, Michael Irwin, Dante Minniti, and HongSheng Zhao for useful
discussions. TMG and WFvA acknowledge partial support from the
National Science Fundation and NASA. SRM was supported by Hubble 
Fellowship Grant Number HF-1036.01-92A awarded to the Space Telescope
Science Institute which is operated by the Association of Universities
for Research in Astronomy, Inc. for NASA under Contract No. NAS5-26555.


\begin{references}
\reference Blanco, V. M. and Blanco, B. M., 1985, Mem. Soc. Astron.
Ital., 56, 15
\reference Carney, B. W., Latham, D. W., Laird, J. B., 1990, \aj, 99, 572
\reference Eggen, O. J., Lynden-Bell, D., and Sandage, A. R.  \apj,  136, 748
\reference Lee, J. -F., and van Altena, W. F., 1983, \aj, 88, 1683
\reference M\'endez, R. A., and van Altena, W. F., 1996, \aj, submitted
\reference Minniti, D., 1993, Ph.D. Thesis, University of Arizona
\reference Minniti, D., 1996, \apj, 459, 175
\reference Plaut, L., 1970, \aap, 8, 341
\reference Plaut, L., 1971, \aaps, 4, 75
\reference Rich, R. M., 1990, \apj, 326, 604
\reference Rich, R. M., Terndrup, D. M., and Sadler, E. M., 1996, in preparation
\reference Spaenhauer, A., Jones, B. F., Whitford, E., 1992, \aj, 103, 297
\reference Tiede, G. P., and Terndrup, D. M., 1996, in preparation
\reference van den Bergh, S., and Herbst, E., 1974, \aj, 79, 603
\reference Zhao, H. S., Spergel, D. N., and Rich, R. M. 1994, \aj,
108, 2154
\reference Zhao, H. S., 1996, M.N.R.A.S., submitted
\reference Zhao, H. S., Rich, R. M., and Biello, J., 1996, \apj, in press
\end{references}
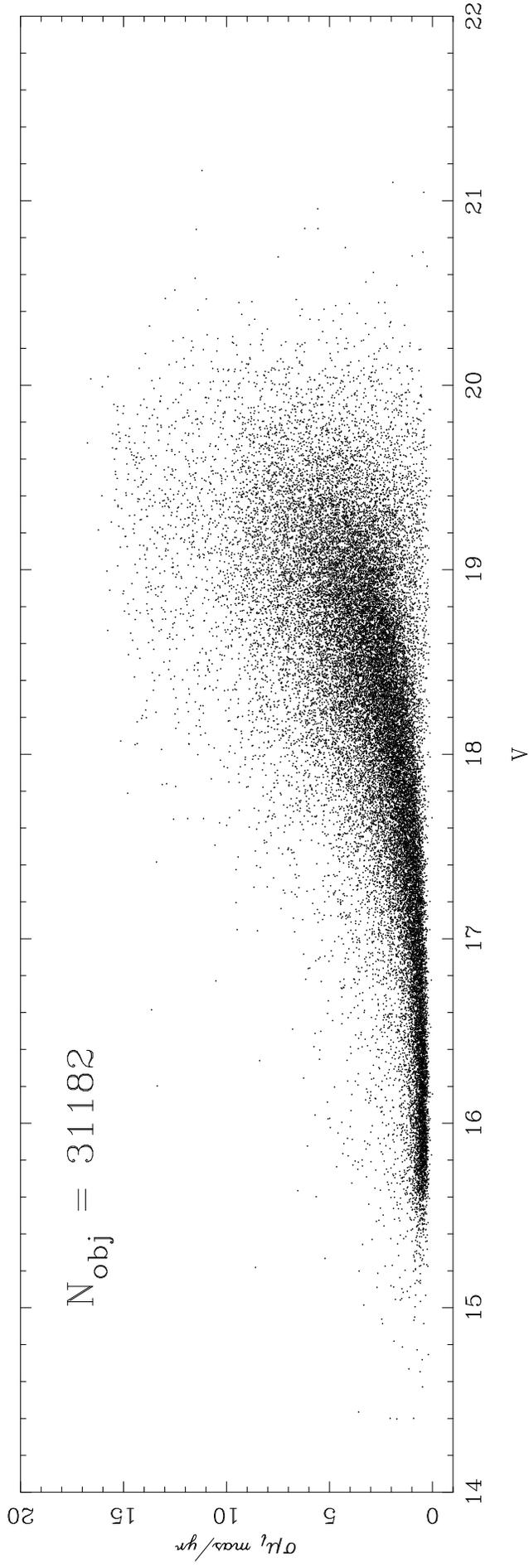
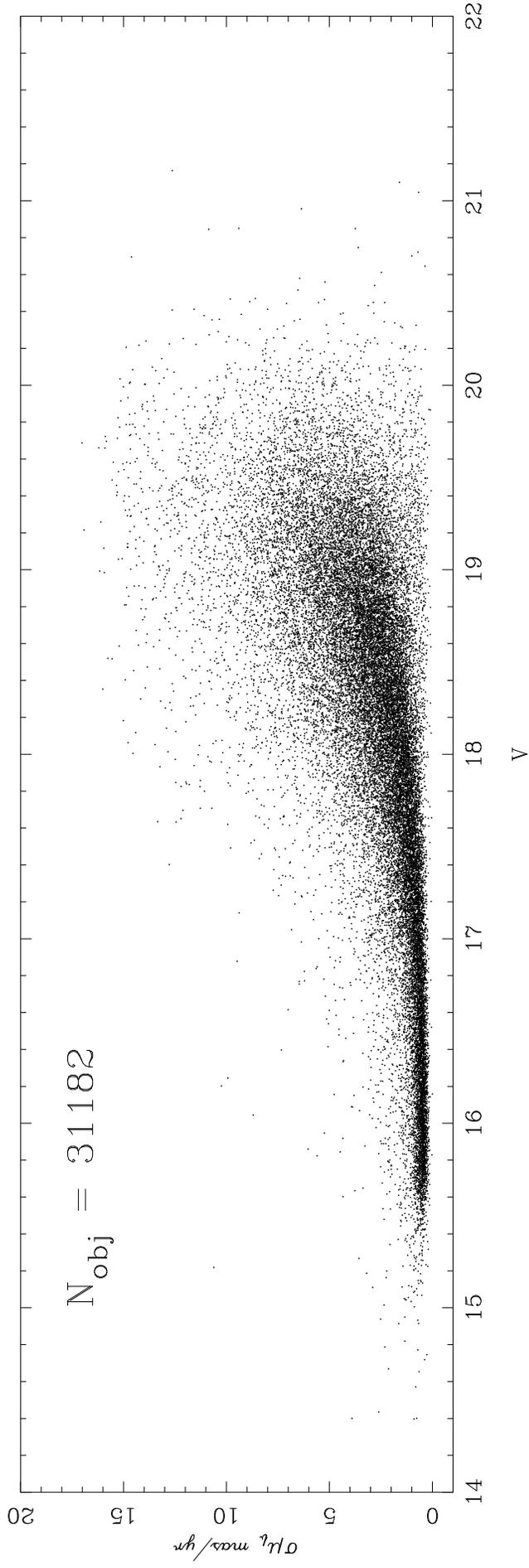
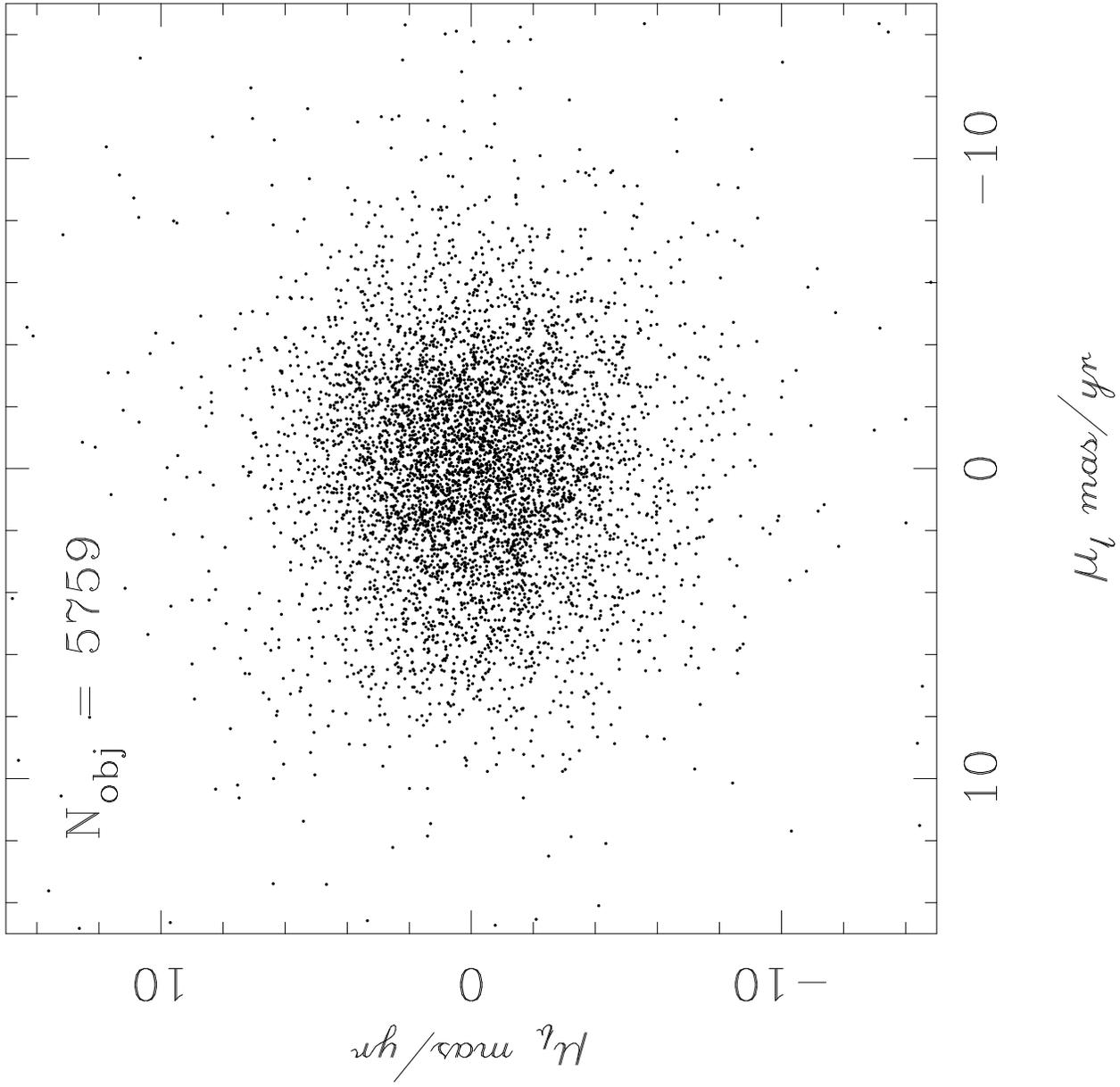
\end{document}